\input harvmac.tex

\input epsf
\ifx\epsfbox\UnDeFiNeD\message{(NO epsf.tex, FIGURES WILL BE IGNORED)}
\def\figin#1{\vskip2in}
\else\message{(FIGURES WILL BE INCLUDED)}\def\figin#1{#1}\fi
\def\ifig#1#2#3{\xdef#1{fig.~\the\figno}
\goodbreak\midinsert\figin{\centerline{#3}}%
\smallskip\centerline{\vbox{\baselineskip12pt
\advance\hsize by -1truein\noindent{\bf Fig.~\the\figno:} #2}}
\bigskip\endinsert\global\advance\figno by1}

\rightline{\vbox{\hbox{CERN-TH/97-175}\hbox{CU-TP-852}
\hbox{\tt hep-th/9707251}}}
\vskip -10 mm
\Title{}{Four-dimensional BPS-spectra via $M$-theory}
\centerline{M{\aa}ns Henningson}
{\it \centerline{Theory Division, CERN}
\centerline{CH-1211 Geneva 23, Switzerland} 
\centerline{\tt henning@nxth04.cern.ch}}
\vskip 5 mm
\centerline{Piljin Yi}
{\it \centerline{Physics Department, Columbia University}  
\centerline{New York, NY 10027, USA}  
\centerline{\tt piljin@phys.columbia.edu}}

\vskip 5mm \centerline{\bf Abstract} 
We consider the realization of four-dimensional theories 
with $N = 2$ supersymmetry as $M$-theory configurations 
including a five-brane. Our emphasis is on the spectrum of massive
states, that are realized as two-branes ending on the five-brane.
We start with a determination of the supersymmetries that are left
unbroken by the background metric and five-brane. We then show how 
the central charge of the $N = 2$ algebra arises from the central charge
associated with the $M$-theory two-brane. This determines the condition 
for a two-brane configuration to be BPS-saturated in the four-dimensional
sense. By imposing certain conditions on the moduli, we can give concrete
examples of such two-branes. This leads us to conjecture that
vectormultiplet and hypermultiplet BPS-saturated states correspond to
two-branes with the topology of a cylinder and a disc respectively.
We also discuss the phenomenon of marginal stability of BPS-saturated
states.
\vskip 5 mm
\Date{\vbox{\line{CERN-TH/97-175\hfill} \line{July 1997 \hfill}}} 

\newsec{Introduction}
Many interesting results about supersymmetric theories have been  
obtained by considering configurations of branes in higher dimensional  
theories. In particular, Witten 
\ref\Witten{
E. Witten, `Solutions of four-dimensional field theory via $M$-theory', 
{\tt hep-th/9703166}.
} 
has recently shown how four-dimensional theories
with $N = 2$ extended supersymmetry can be realized as $M$-theory  
configurations including a supersymmetric five-brane.
In many cases, this construction gives an easy way to determine the  
spectral
curve and the associated meromorphic one-form that appears in the
Seiberg-Witten formulation 
\ref\Seiberg-Witten{
N. Seiberg and E. Witten, `Electric-magnetic duality, monopole  
condensation,
and confinement in $N = 2$ supersymmetric Yang-Mills theory', 
{\it Nucl. Phys.} {\bf B426} (1994) 19, {\tt hep-th/9407087} \semi
N. Seiberg and E. Witten, `Monopoles, duality and chiral symmetry  
breaking
in $N = 2$ supersymmetric QCD', {\it Nucl. Phys.} {\bf B431} (1994)  
484, {\tt hep-th/9408099}.
}
of $N = 2$ supersymmetric gauge theories.

Our focus in this paper is on the spectrum of massive BPS-saturated states
in such theories. These states enjoy a particular stability property, that 
ensures that they can only decay as certain so called marginal  stability
domain walls of codimension one in the moduli space of vacua are crossed.
Given the existence of a BPS-saturated state with certain quantum  
numbers, Seiberg-Witten theory gives a formula for its contribution to the
central charge of the $N = 2$ algebra and thus for its mass. However, it is
in general a difficult problem to determine the spectrum of such states that
really exist at a given point in the moduli space.
(See for example
\ref\Bilal{
A. Bilal and F. Ferrari, `The strong-coupling spectrum of the Seiberg-Witten
theory', {\it Nucl. Phys.} {\bf B469} (1996) 387, {\tt hep-th/9602082}.
}
for the case of the pure $N = 2$ $SU(2)$ Yang-Mills theory.) 
The $M$-theory interpretation has the conceptual advantage of 
giving a fairly concrete picture of such states: 
They correspond to two-branes with boundary on the five-brane. 
The homology class of the boundary determines the quantum numbers of the 
state, and BPS-saturation amounts to the world-volume being minimal in
an appropriate sense. 
The problem of finding the BPS-saturated spectrum is therefore in 
principle reduced to the problem of finding such minimal world-volumes.
(Another approach, that amounts to finding geodesic curves on the five-brane,
was initiated in
\ref\Klemm{
A. Klemm, W. Lerche, P. Mayr, C. Vafa and N. Warner,
`Self-dual strings and $N = 2$ supersymmetric field theory',
{\it Nucl. Phys.} {\bf B477} (1996) 746, {\tt hep-th/9604034}.
}.) 

This paper is organized as follows: 
In section two, we determine the
unbroken supersymmetries in these $M$-theory configurations. The 
eleven-dimensional manifold on which the theory is defined must contain a 
four-dimensional hyper K\"ahler manifold as a factor. 
The five-brane configuration
is described by a two-dimensional surface in this space that
is holomorphically embedded with respect to one of its complex structures.
In section three, we show how the eleven-dimensional central charge 
associated with a
two-brane gives rise to the central charge of the $N = 2$ algebra in four dimensions, and determine the conditions for a two-brane to be BPS-saturated.
It turns out that the two-brane must also be described by a two-dimensional
surface which is holomorphically embedded, but with respect to another
complex structure which is orthogonal to the original one. The remaining
freedom in choosing this second complex structure corresponds to the phase
of the central charge of the BPS-saturated state. 
In section four, we give some concrete examples of BPS-saturated states by
imposing certain reality conditions on the moduli. These examples lead to 
the conjecture that BPS-saturated vectormultiplets and
hypermultiplets correspond to surfaces with the topology of a cylinder and a
disc respectively. We also discuss the phenomenon of decay of BPS-saturated
states across domain walls of marginal stability and its relation to
the mutual non-locality of the constituent states.

Throughout this paper, we exemplify the general discussion by considering
the case of the low-energy effective theory of $N = 2$ QCD, i.e. $SU(N_c)$
super Yang-Mills theory with some number $N_f$ of hypermultiplet quarks
in the fundamental representation.

While preparing this manuscript, we were informed that closely related
issues would be addressed in
\ref\Princeton{A. Mikhailov, `BPS-states and minimal surfaces',
{\tt hep-th/9708068}.
}.

\newsec{$N = 2$ supersymmetry in $d = 4$ dimensions from $M$-theory}
Following \Witten, we consider $M$-theory on an eleven-dimensional 
manifold $M^{1, 10}$ of the form
\eqn\Meleven{
M^{1, 10} \simeq {\bf R}^{1, 3} \times X^7 ,
}
where the first factor is the four-dimensional Minkowski space-time and  
the second factor is some (non-compact) seven-manifold. Furthermore, we  
introduce a five-brane, the world-volume of which fills ${\bf R}^{1,  
3}$ and defines a two-dimensional surface $\Sigma$ in $X^7$.  

The supercharges $Q_A$, $A = 1, \ldots, 32$ of $M$-theory transform as  
a Majorana spinor under the eleven-dimensional Lorentz group. In a flat  
space background they fulfill the supertranslations algebra
\ref\Townsend{
P. K. Townsend, `P-brane democracy', {\tt hep-th/9507048}.
}
\eqn\Msusy{
\{ Q_A, \bar{Q}_B \} = (\Gamma^M)_{AB} P_M + (\Gamma^{M N})_{AB}  
Z_{M N} + (\Gamma^{M N P Q R})_{AB} W_{M N P Q R} .
}
Here $P_M$ is the eleven-dimensional  
energy-momentum ($M, N, \ldots = 0, 1, \ldots, 10$), 
and the central charges $Z_{M N}$ and $W_{M N P Q R}$ 
are associated with the two-brane and the five-brane respectively:
\eqn\charges{
\eqalign{
Z^{M N} & \sim \int_{2-brane} d X^{M} \wedge d X^{N} \cr
W^{M N P Q R} & \sim \int_{5-brane} d X^M \wedge \ldots \wedge  
d X^R . \cr
}
}
Part of the eleven-dimensional supersymmetry generated by the $Q_A$
is spontaneously broken by the background configuration:
An unbroken supersymmetry generator must  
first of all be covariantly constant with respect to the background  
metric on $M^{1, 10}$. 
Furthermore, it follows from the algebra \Msusy\ and the  
expression \charges\ for the central charges together with the fact  
that the mass of a brane is proportional to its area, that an unbroken  
supersymmetry generator must be of positive chirality with respect to  
the tangent-space of each five-brane world-volume element. In the
rest of this section, we will analyze the implications of these
requirements in more detail. 

\subsec{The background metric}
We are interested in background configurations with an unbroken $N = 2$  
extended supersymmetry in ${\bf R}^{1, 3}$.
To this end, we first require the holonomy group of $X^7$ to be isomorphic 
to $SU(2)$ 
(rather than to $SO(7)$ as it would be for a generic seven-manifold). 
The spinor representation of $SO(7)$ then
contains four singlets with respect to the $SU(2)$ holonomy 
group, so $X^7$ admits four covariantly constant spinors. The holonomy 
of the background metric on $X^7$ thus leaves an $N = 4$ extended 
supersymmetry in ${\bf R}^{1, 3}$ unbroken. 
Furthermore, the vector  
representation of $SO(7)$ contains three singlets with respect to the  
$SU(2)$ holonomy group. This means that $X^7$ must be of the form
\eqn\Xseven{
X^7 \simeq {\bf R}^3 \times Q^4 ,
}
where ${\bf R}^3$ is generated by the flows of the three covariantly  
constant vector fields on $X^7$, and $Q^4$ is a four-manifold of  
$SU(2)$ holonomy, i.e. a hyper-K\"ahler manifold. 
The 4 covariantly constant spinors of $X^7$
transform as a pair of doublets under the $SO(3)\simeq SU(2) / {\bf  
Z}_2$ Lorentz group of ${\bf R}^3$. Each doublet corresponds to one of  
the two covariantly constant Majorana spinors on $Q^4$, which we will  
denote by $\zeta_1$ and $\zeta_2$. 

Hyper-K\"ahler geometry can be thought of as a K\"ahler geometry which
admits a family of inequivalent complex structures parametrized by a
two-sphere $S^2$. In four dimensions, the hyper-K\"ahler condition 
is in fact equivalent to the Calabi-Yau condition, i.e. $Q^4$ is a
Ricci-flat K\"ahler manifold and should therefore admit a covariantly
constant holomorphic two-form $\Omega$. Given a choice
of complex structure $J$, i.e. a point on the two-sphere $S^2$,
$\Omega$ is uniquely determined 
(up to a complex constant) as $K^\prime + i K^{\prime \prime}$, 
where $K^\prime$ and $K^{\prime \prime}$ are the K\"ahler forms 
corresponding to two other
complex structures $J^\prime$ and $J^{\prime \prime}$ that are orthogonal
to each other and also to the chosen complex structure $J$.   

\subsec{The five-brane geometry}
Given the background metric described in the previous subsection, we will
now determine the constraints on the embedding of a five-brane for the
resulting configuration to leave $N = 2$ supersymmetry unbroken in
${\bf R}^{1, 3}$. First of all, unbroken Poincar\'e invariance requires  
that
the five-brane world-volume fills all of ${\bf R}^{1, 3}$ and defines a
two-dimensional surface $\Sigma$ in $X^7$. Furthermore, we will later 
identify (the double cover of) the Lorentz group $SO(3)$ of the first 
factor of \Xseven\ with the $R$-symmetry of the $N = 2$ algebra. To  
leave this unbroken, $\Sigma$ must in fact lie at a single point in ${\bf  
R}^3$ and define a surface in $Q^4$, that by a slight abuse of notation 
we also call $\Sigma$.

A generic five-brane configuration, described by a surface $\Sigma$ in $Q^4$,
will break all supersymmetries. To have $N = 2$ supersymmetry unbroken,
we must require the spatial volume element of the surface $\Sigma$
to be minimized so that it saturates a topological bound. Given a choice
of complex structure $J$ on $Q^4$, 
a useful identity in this regard involves the pull-backs $K_\Sigma$
and $\Omega_\Sigma$ of the corresponding K\"ahler form $K$ 
and the holomorphic two-form $\Omega$ to $\Sigma$.
A straightforward computation shows that
\eqn\identity{
{1 \over 4} \left(({}^*K_\Sigma)^2 + |{}^*\Omega_\Sigma|^2\right) = 1 ,
}
where ${}^*$ denotes the Hodge-dual with respect to the induced metric
on $\Sigma$. It follows that the area $A_\Sigma$ of $\Sigma$ fulfills
\eqn\Asigma{
2 A_\Sigma = 2 \int_\Sigma V_\Sigma
= \int_\Sigma V_\Sigma \sqrt{({}^*K_\Sigma)^2 + |{}^*\Omega_\Sigma|^2}
\geq \int_\Sigma V_\Sigma {}^*K_\Sigma = \int_\Sigma K_\Sigma ,
}
were $V_\Sigma$ denotes the volume-form of $\Sigma$ and the inequality
is saturated if and only if $\Omega_\Sigma$ vanishes identically.
This is equivalent to requiring $\Sigma$ to be holomorphically embedded in $Q^4$ with respect to the complex structure
$J$. The surface $\Sigma$ is thus the locus of the equation
\eqn\Sigmasurface{
F = 0
}
for some holomorphic function $F$ on $Q^4$\footnote{\dag}{When we
are dealing with an infinite surface $\Sigma$, the lower bound
for the total area does not make much sense. One should regard it as a  
local statement in that the minimization is with respect to all possible 
localized perturbation of the surface.}.
We emphasise that the choice
of the complex structure $J$ is arbitrary.  

We will now determine the  
amount of supersymmetry that is left unbroken by this configuration. 
Recall that the unbroken supersymmetry generators should have
positive chirality with respect to each five-brane
world-volume element. Since the five-brane world-volume is of the form
${\bf R}^{1, 3} \times \Sigma$, we see that this requirement correlates
space-time chirality and chirality with respect to 
each $\Sigma$ area element and thus breaks at least half the  supersymmetry.
In fact, there exist certain linear combinations 
$\zeta^+$ and $\zeta^-$ of the covariantly constant Majorana spinors
$\zeta_1$ and $\zeta_2$ on $Q^4$ with {\it constant} coefficients
that have definite and opposite chirality
with respect to every area element of $\Sigma$, 
regardless of the form of the holomorphic function $F$.
The five-brane thus breaks 
exactly half of the $N=4$ supersymmetry, and we are left with $N = 2$
supersymmetry in ${\bf R}^{1, 3}$.
 
To verify the existence of such $\zeta^+$ and $\zeta^-$, 
we consider the spinor representation 
$4_s = (2, 1) \oplus (1, 2)$ of the rotation group  
$SO(4) \simeq SU(2)_L \times SU(2)_R$ of $Q^4$.
This representation decomposes into $2_0 \oplus 1_{-1/2} \oplus  
1_{+1/2}$ 
under the subgroup $U(2) \simeq  SU(2)_L \times U(1)_R$, 
where the $SU(2)_L$ factor is the holonomy group of $Q^4$ and 
$U(1)_R$ is the group of rotations in the tangent 
space of the holomorphically embedded surface $\Sigma$. The two 
spinors $\zeta^\pm$ above correspond to the two singlets of $SU(2)_L$ 
and can thus be chosen to be chiral eigenstates of opposite sign 
with respect to an area-element of $\Sigma$. 
But the form of $\zeta^\pm$ depends only on the choice of $U(1)_R$ in
$SU(2)_R$, tantamount to the choice of the complex structure,
so $\zeta^\pm$ are indeed constant linear combinations of 
$\zeta_1$ and $\zeta_2$ as claimed. 

\subsec{The case of $N = 2$ QCD}
We will exemplify the general discussion above with the case of the
low-energy effective theory of $SU(N_c)$ super Yang-Mills theory with
some number $N_f$ of quark hypermultiplets in the fundamental
representation. As usual, we will impose  
the restriction that $0 \leq N_f < 2 N_c$ so that the theory is  
asymptotically free. 

In this case, $Q^4$ is ${\bf R}^3 \times S^1$ with the standard flat
metric \Witten. 
We introduce coordinates $X^4$, $X^5$, $X^6$ and $X^{10}$ on $Q^4$, 
where $X^{10}$ is periodic with period $2 \pi R$ for some $R$. 
In the following we will set $R=1$.
We choose a particular complex structure $J$ such that $s = X^6 + 
i X^{10} $ and  $v = X^4 + i X^5$ are holomorphic coordinates. 
The corresponding K\"ahler form is given by
\eqn\Kahlerform{
K = i (d s \wedge d \bar{s} + d v \wedge d \bar{v} ) ,
}
and the covariantly constant holomorphic two-form is
\eqn\Omegaform{
\Omega = d s \wedge d v .
}
It is convenient to replace $s$ by the single-valued coordinate 
$t = \exp (-s)$.

Such a $Q^4$ obviously has a trivial holonomy group and admits
two additional covariantly constant Majorana spinors $\zeta_3$ and 
$\zeta_4$. The background metric on $Q^4$ thus leaves $N = 8$ supersymmetry
unbroken in ${\bf R}^{1, 3}$ rather than $N = 4$, but the five-brane
again breaks the symmetry down to $N = 2$. The reason is that although
it is possible to form linear combinations $\tilde{\zeta}^+$ and
$\tilde{\zeta}^-$ of $\zeta_3$ and $\zeta_4$ of definite
and opposite chirality with respect to a given area element of $\Sigma$,
the coefficients are no longer constant but will vary over $\Sigma$.
The spinors $\tilde{\zeta}^+$ and $\tilde{\zeta}^-$ are thus not 
covariantly constant and do not give rise to unbroken supersymmetries
in ${\bf R}^{1, 3}$. 

The Riemann surface $\Sigma$ that describes the five-brane world-volume
is given by the equation 
\eqn\curve{
F(t,v)\equiv t^2 + B(v) t + C(v) = 0 ,
}
where
\eqn\BC{
\eqalign{
B(v) & = \prod_{a = 1}^{N_c} (v - \phi_a) \cr
C(v) & = \Lambda^{2 N_c - N_f} \prod_{i = 1}^{N_f} (v - m_i) . \cr
}
}
Here the $m_i$, $i = 1, \ldots, N_f$ are the bare masses of the 
hypermultiplet quarks, the  
$\phi_a$, $a = 1, \ldots, N_c$ parametrize the moduli space of vacua 
subject to the constraint
\eqn\constraint{
\sum_{a = 1}^{N_c} \phi_a = 0 ,
}
and $\Lambda$ is the dynamically generated scale of the theory.
This form of $\Sigma$ was first proposed in
\ref\Hanany{
A. Hanany and Y. Oz, `On the quantum moduli space of vacua of $N = 2$
supersymmetric $SU(N_c)$ gauge theories', {\it Nucl. Phys.} {\bf B452}
(1995) 283, {\tt hep-th/9505075}.
}\ref\Argyres{
P.C. Argyres, M. R. Plesser and A. D. Shapere, `The Coulomb phase of
$N = 2$ supersymmetric QCD', {\it Phys. Rev. Lett.} {\bf 75} (1995) 1699,
{\tt hep-th/9505100} .
}.
If we parametrize the surface $F=0$ by $v$, we necessarily encounter 
singular points because $v$ spans a complex plane or a $CP^1$, while  
the actual surface has genus $N_c - 1$. The solutions to the equation $F=0$ 
for a given $v$, i.e. 
\eqn\doublecover{ 
t = t_\pm(v)= -{B(v)\over 2}\pm\sqrt{ \left({B(v)\over 2}\right)^2-C(v)}
}
gives the holomorphic embedding of $\Sigma$ as a double cover over  
$v$-plane. In the regime of not
too strong coupling, i.e. when $|\Lambda|$ is small compared to the
distances between the $\phi_a$ and the $m_i$, we see that there is a pair
of branch points, that we join by a branch cut, in the vicinity of each
$\phi_a$. Note that the geometries of the two sheets are not identical. 
In particular, the upper sheet given by $t_+(v)$ goes off to $X^6=+\infty$ 
(i.e. $t_+\rightarrow 0$) whenever $C(v)$ vanishes, i.e. at
$v = m_i$, $i = 1, \ldots, N_f$. See figure 1 below.

\ifig\sheet{The surface $\Sigma$ as a double cover of the $v$-plane for
the case of $N_c = 5$ and $N_f = 1$. Five pairs of branch points (solid
dots) are joined by branch cuts (wavy lines). The quark singularity 
on the upper sheet is
represented by a hollow dot. We have also indicated representatives of
three different homology cycles on $\Sigma$. 
Dashed and dotted contours are on the
lower and upper sheet respectively. Note that the homology cycle denoted
$W$ is represented by two closed curves of opposite orientation on the
upper sheet.}
{\epsfysize 3in\epsfbox{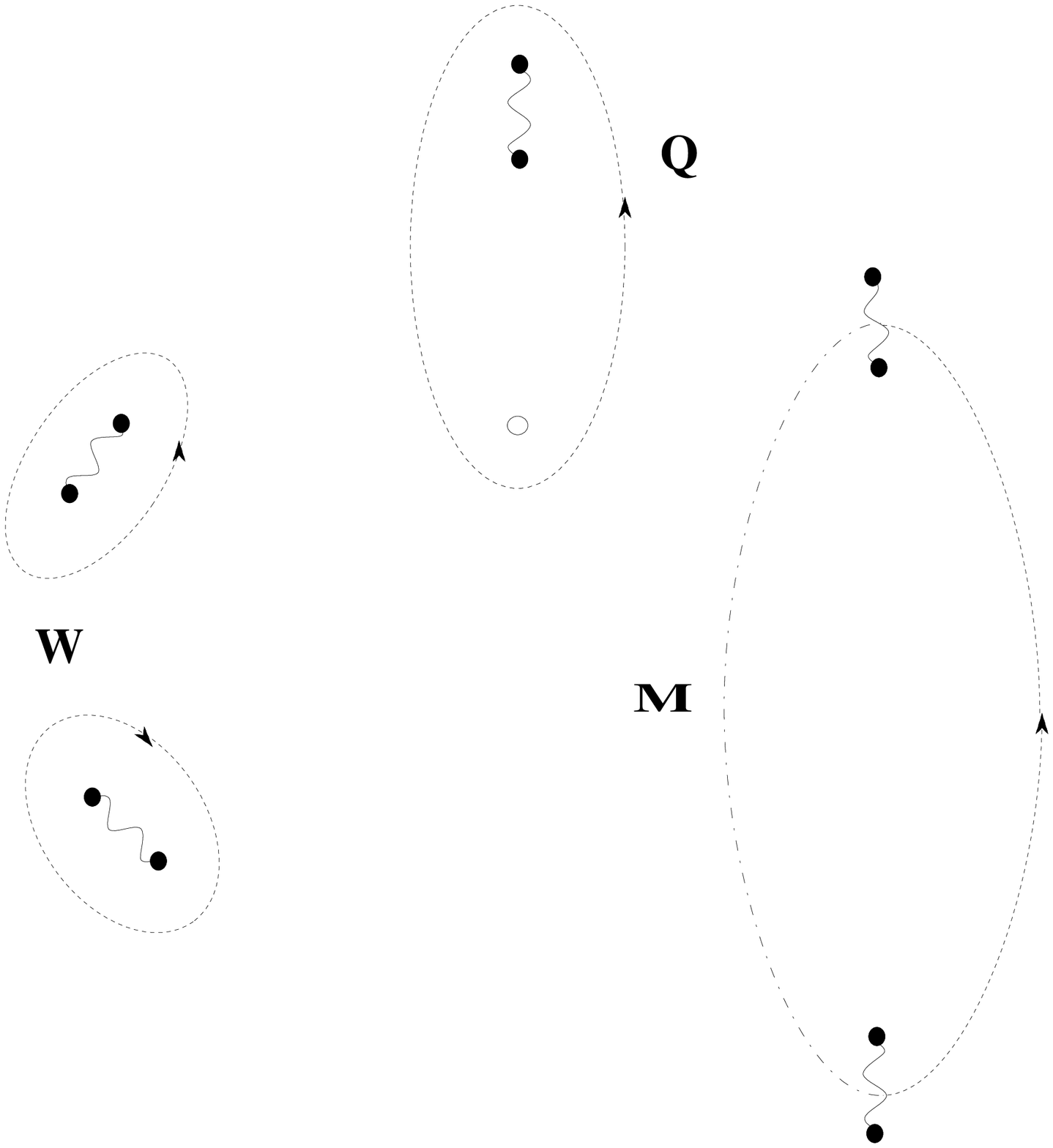}}

\newsec{Supersymmetric two-branes}
We will now consider the BPS-saturated excitations of the vacuum state
defined by the background metric and five-brane configuration described  
in the previous subsections. These states are realized by
supersymmetric configurations of two-branes. Since $\Sigma$ plays 
the role of the spectral curve in the Seiberg-Witten formalism, 
the two-brane must have boundaries on
$\Sigma$ to constitute a charged state. Since $\Sigma$ lies 
at a single point in ${\bf R}^3$, it is reasonable that this should hold 
for the two-brane as well. 
That is, we now consider a two-brane  excitation, the 
world-volume of which spans a world-line in ${\bf  R}^{1, 3}$, lies 
at a single point in ${\bf R}^3$ and defines a  two-dimensional surface 
$D$ in $Q^4$. The boundary $C = \partial D$ of  $D$ lies on $\Sigma$ so 
that the two-brane ends on the five-brane 
\ref\Strominger{
A. Strominger, `Open $p$-branes', {\it Phys. Lett.} {\bf B383} (1996) 44,
{\tt hep-th/9512059}.
}.

The general form of the lower-bound for the area of a brane
found in the previous 
section equally applies for the two-brane. The area is thus again bounded 
from below by a topological quantity given as the integral of the pull-back 
to the two-brane surface $D$ of the K\"ahler-form corresponding to
some complex structure $J^\prime$. Suppose now that
the bound is saturated, so that $D$ is holomorphically embedded with 
respect to $J^\prime$. The complex structure $J^\prime$ cannot equal the
complex structure $J$ with respect to which the surface $\Sigma$ is
holomorphically embedded, because 
this would mean that the intersection $C$ of $D$ and $\Sigma$ would also
be a holomorphically embedded submanifold of $Q^4$ and thus in particular
could not be a curve of real dimension one.
An obvious guess is then to require that $J^\prime$ should be orthogonal
to $J$.
Given $J$, the set of such $J^\prime$ is parametrized by an $S^1$, that
would naturally correspond to the phase of the central charge of the
BPS-saturated state.

To see this more explicitly, let us construct the unbroken  
supersymmetry generators in the presence of the five-brane: 
We decompose an eleven-dimensional vector, such as the energy-momentum
$P_M$, $M = 0, \ldots, 10$ into a vector $P_\mu$, $\mu = 0, 
\ldots, 3$ in ${\bf R}^{1, 3}$, a vector $P_m$, $m = 7, 8, 9$ in ${\bf  
R}^3$ and a vector  $P_s$, $s = 4, 5, 6, 10$ in $Q^4$. An eleven-dimensional  
spinor, such  as the supercharges $Q_A$, 
$A = 1, \ldots, 32$ is written as $Q_{\alpha a i}$,  
where  $\alpha = 1, \ldots, 4$, $a = 1, 2$ and $i = 1, \ldots, 4$ index a  
spinor on ${\bf R}^{1, 3}$, ${\bf R}^3$ and $Q^4$ respectively. The  
supercharges that are left unbroken by the background metric and  
five-brane configuration described above can be written schematically as
\eqn\etaspinor{
\eta_{\alpha a} = \Pi^+_{\alpha \beta} \zeta^+_i Q_{\beta a i} +  
\Pi^-_{\alpha \beta} \epsilon_{a b} \zeta^-_i Q_{\beta b i} ,
}
where $\Pi^\pm$ are the projectors on positive and negative ${\bf  
R}^{1, 3}$ chirality, $\epsilon_{ab}$ is anti-symmetric and $\zeta^\pm$
are the covariantly constant spinors on $Q^4$ that we discussed above. 
This particular expression is valid to the extent that 
the $\zeta^\pm$ are uniform in the local coordinates chosen.
More generally, one must actually include $\zeta^\pm$ inside the 
spatial integration that defines the global supersymmetry generators. 
The $\eta_{\alpha a}$ in \etaspinor are constructed so that  
they are Majorana spinors on ${\bf R}^{1, 3}$. To see this, recall that
the supercharges $Q_A$ form a Majorana spinor on ${\bf  
R}^{1, 10}$. Furthermore, the charge conjugation matrix on ${\bf R}^{1,  
10}$ is the tensor product of the charge conjugation matrices on ${\bf  
R}^{1, 3}$,  ${\bf R}^3$ and $Q^4$, the second of which equals the 
antisymmetric  tensor $\epsilon_{a b}$. Finally, the covariantly constant 
spinors $\zeta^+$ and $\zeta^-$ are each others charge conjugates on $Q^4$.

Inserting the form of $\eta_{\alpha a}$ in the $11$-dimensional  
supersymmetry algebra \Msusy, we find that they fulfill
\eqn\etaalgebra{
\{ \eta_{\alpha a}, \bar{\eta}_{\beta b} \} = (\gamma^\mu)_{\alpha  
\beta} P_\mu \delta_{a b} + \epsilon_{a b} \left( \Pi^+_{\alpha \beta}  
Z + \Pi^-_{\alpha \beta} \bar{Z} \right) .
}
Here we have normalized $\zeta^+$ and $\zeta^-$ by $\bar{\zeta}^+  
\zeta^+ = \bar{\zeta}^- \zeta^- = 1$, and the central charge $Z$ is  
given by
\eqn\Zcharge{
Z = \int_{D} \Omega_D ,
}
where $\Omega_D$ is the pull-back to $D$ of the covariantly constant 
two-form $\bar{\zeta}^- \gamma_{s t} \zeta^+ d X^s 
\wedge d X^t$ on $Q^4$. As the notation suggests and as a little algebra shows,
the latter two-form
in fact equals the covariantly constant {\it holomorphic} $(2,0)$ 
form $\Omega$ of $Q^4$ that we described in the last section.

We can now use an identity analogous to \identity:
\eqn\identitytwo{
{1 \over 4} \left(({}^*K_D)^2 + |{}^*\Omega_D|^2\right) = 1 ,
}
where $K_D$ is the pull-back of the K\"ahler form $K$ to $D$, 
and the ${}^*$ denotes the Hodge-dual with
respect to the induced metric on $D$.
It follows that the area $A_D$ of $D$ fulfills the
inequalities in 
\eqn\bps{
2 A_D = 2 \int_D V_D= \int_D V_D \sqrt{({}^*K_D)^2 + |{}^*\Omega_D|^2} 
\geq \int_D \left|{}^*\Omega_D \right| V_D \geq 
\left| \int_D \Omega_D \right| = \left| Z  
\right| ,
}
where $V_D$ is the volume-form of $D$.
The first  
inequality is saturated if and only if the pull-back $K_D$ of the
K\"ahler form vanishes, 
while the second is 
saturated if and only if the phase of the pull-back ${}^*\Omega_D$ 
of the holomorphic two-form is constant over $D$. 
This constant phase of ${}^*\Omega_D$ tells us that there is a second
K\"ahler form whose pull-back to $D$ vanishes. The surface $D$ is then
a holomorphic embedding with respect to some complex structure $J^\prime$ 
which is orthogonal to the complex structure $J$.
Given $J$, there is an $S^1$ of possible such 
$J^\prime$, corresponding to the  
phase of ${}^*\Omega_D$, i.e. the phase of the central charge $Z$. Different 
supersymmetric surfaces $D$ will in general have different constant 
phases of ${}^*\Omega_D$, and will be holomorphic with respect to 
different $J^\prime$.
 
Assuming that $\Omega_D$ is exact on $D$, i.e. $\Omega_D = d\lambda$ for
some one-form $\lambda$, 
the central charge can be rewritten as a boundary  
integral,
\eqn\Omegaexact{
Z=\int_D \Omega_D=\oint_C \lambda .
}
This defines a one-form $\lambda$ on $\Sigma$ up to an exact holomorphic 
form.  This one-form on $\Sigma$ is closed because $d\lambda$ is the  
pull-back of the holomorphic two-form $\Omega$ to holomorphically embedded
surface $\Sigma$, and must vanish identically. 
We have thus recovered 
the Seiberg-Witten expression for the central charge as the period of 
a certain  meromorphic one-form $\lambda$ on the spectral curve $\Sigma$.  

\subsec{The case of $N = 2$ QCD}
Coming back to the example of $SU(N_c)$ super Yang-Mills theory
with $N_f$ flavors, we can see
that the holomorphic two-form $\Omega=ds\wedge dv$ is indeed exact in this
case. A possible choice for $\lambda$ is
\eqn\oneform{
\lambda=-v\,ds_\pm(v)=v\,{dt_\pm(v)\over t_\pm(v)},
}
where $t_\pm(v)$ solve the equation $t^2+B(v)t+C(v)=0$. This form of
$\lambda$ was also noted in
\ref\Landsteiner{
K. Landsteiner, E. Lopez and D. A. Lowe, `$N = 2$ supersymmetric gauge
theories, branes and orientifolds', {\tt hep-th/9705199}.
}\ref\Fayyazuddin{
A. Fayyazuddin and M. Spali\'nski, `The Seiberg-Witten differential 
from $M$-theory', {\tt hep-th/9706087}.
}. 
More explicitly, one finds
\eqn\pullbackoneform{
\lambda= -v dv {1 \over 2 t_\pm (v) + B(v)}\left( B^\prime (v)  +
{C^\prime (v) \over  t_\pm(v)} \right) .
}
Note that $\lambda$ possesses a pole on the upper sheet when $t_+(v) =  
0$, which happens whenever $C(v)=0$, i.e., at $v = m_i$. This pole is 
associated with the bare mass of a quark and has residue equal to $m_i$.

In this example, it is also easy to construct the complex structures that
are orthogonal to the one with respect to which $s$ and $v$ are holomorphic.
Such a complex structure is specified by an angle $\theta$ and can be
described by stating that the variables
\eqn\zandw{
\eqalign{
z & = {1 \over 2} 
\left( s + \bar{s} + e^{i \theta} v - e^{-i \theta} \bar{v} \right) \cr
w & = {1 \over 2} 
\left( -s + \bar{s} + e^{i \theta} v + e^{-i \theta} \bar{v} \right)
}
}
are holomorphic. In these variables, the original K\"ahler form and the holomorphic two-form \Kahlerform\ and \Omegaform\ are
\eqn\Kform{
\eqalign{
K & = i (d z \wedge d w - d \bar{z} \wedge d \bar{w}) \cr
\Omega & = {1 \over 2} e^{-i \theta} 
(-d z \wedge d \bar{z} - d w \wedge d \bar{w} + d z \wedge d w
+ d \bar{z} \wedge d \bar{w} ) ,
}
}
so $K$ indeed vanishes and $\Omega$ has a constant phase determined by
$\theta$ when pulled back to a surface $D$ which is holomorphically embedded
with respect to this complex structure.

\newsec{The BPS-saturated spectrum}
In this section, we will discuss the spectrum of BPS-saturated states  
in general $N = 2$ theories. Although Seiberg-Witten theory gives a way of
calculating the central charge, and thus the mass, of a BPS-saturated  
state with given quantum numbers, it does not address the question of whether  
such a state really exists in the spectrum. The $M$-theory approach in  
principle provides an answer; the existence of a state amounts to the 
existence of a supersymmetric surface $D$, the boundary of which lies on the  
spectral curve $\Sigma$ and represents a specific homology class given by the  
quantum numbers of the state. As we have seen, such a surface is
holomorphically embedded with respect to a complex structure that is
given by the phase of the central charge.

It is important to emphasize that the BPS condition requires more 
than the simple minimization of the mass. Given a homology 
class of the boundary, there will in general exist surface of minimal  
area, but there is no guarantee that it actually saturates the BPS bound. 
Conversely, a generic supersymmetric surface $D$, i.e. a surface that  
is holomorphically embedded with respect to the complex structure 
$J^\prime$, will only intersect the surface $\Sigma$ 
at isolated points and thus fail to be a finite surface.
Furthermore, even if $D$ intersects $\Sigma$ along a closed curve $C$,  
it may still go off to infinity in $Q^4$ somewhere in the interior.
And here lies the difficulty in finding the actual BPS spectrum.

\subsec{The case of $N = 2$ QCD}
Before proceeding with the more general discussion, we would like
to present some examples of exact supersymmetric two-brane configurations 
in the case of $N = 2$ QCD that we have been discussing. We can describe the
quantum numbers of a state by giving the homology class of the boundary on
$\Sigma$ of the corresponding surface $D$. At not too strong coupling, 
the BPS-saturated spectrum is known to consist of the following states:

\vskip 5mm

1. Electrically charged vector mesons $W$ in vectormultiplets.

2. Magnetically charged monopoles $M$ in hypermultiplets.

3. Electrically charged quarks $Q$, that also carry a quark number charge,
in hypermultiplets.

4. Electrically and magnetically charged dyons $D$ in hypermultiplets.

5. Quark-soliton bound states $B$ in hypermultiplets.

\vskip 5mm

Representatives of the homology classes corresponding to $W$, $M$ and $Q$
were indicated in figure 1 above. Some of these states may decay as they cross
a domain wall of marginal stability. 

We will now give examples of the $D$-surfaces corresponding
to vectormultiplets, monopoles and quarks. We will do this in a limit
where the branch points and quark singularities not relevant for the state
in question are far away and thus can be disregarded. To be able to give
exact solutions, we will also impose a certain reality condition on the moduli.

For the vector mesons and the monopoles, the condition that the irrelevant
features of $\Sigma$ are far away means that we are effectively considering
the pure $SU(2)$ theory, 
i.e. the surface is given by the simple equation
\eqn\simplesigma{
t = t_\pm (v) = - {v^2 - \phi^2 \over 2} \pm 
\sqrt{ \left( {v^2 - \phi^2 \over 2} \right)^2 - {1 \over 4}} ,
}
where $\phi=\phi_1=-\phi_2$ is the complex adjoint Higgs expectation
value and we have set the scale $\Lambda = 1 / \sqrt{2}$. 
The square-root branch cuts emanate from the points 
$v=\pm \sqrt{\phi^2 + 1}$ and end at the points 
$v=\pm \sqrt{\phi^2 - 1}$.

The reality condition that we will impose on the modulus in order to
be able to construct exact BPS-saturated two-branes
amounts to taking $\phi$ to be real and and $|\phi| > 1$ so that all 
the branch points lie on the real $v$-axis. (One could also take $\phi$
to be imaginary and $|\phi| > 1$, in which case all branch points
would lie on the imaginary $v$-axis.)

In the vicinity of one of the branchpoints, say at $v=v_0$, 
the induced metric on $\Sigma$ has the form
\eqn\inducedmetric{
g_\Sigma= \left|{dt_\pm(v)\over t_\pm(v)}\right|^2+|dv|^2\simeq 
|v_0| \left| {dv \over \sqrt{v - v_0}} \right|^2,
}
which means that $v-v_0$ covers only half of the local neighborhood. A  
better
holomorphic coordinate would be $y=\sqrt{v-v_0}$, for which one finds
\eqn\newcoordinate{
g_\Sigma \simeq 4 | v_0 | | dy |^2 .
}
For instance, a closed path on $v$-plane
that tightly winds around a pair of such singular
points $\sqrt{\phi^2 + 1}$ and $\sqrt{\phi^2 - 1}$ is actually
a smooth circle on the surface $\Sigma$, although it more looks like
a pair of straight segments between the two branch points in the 
$v$-plane. See figure 2 below.

\ifig\sheet{A half of the surface $\Sigma$. The first figure shows
the square root cuts (wiggly lines) in the $v$-plane, while the second
illustrates the actual shape of the surface. The second half given by  
the other sheet is smoothly attached to the two circles to the right. As  
one tightens the closed paths around the branch cuts in (a), the  
corresponding circles in (b) move toward the other half of $\Sigma$. 
}{\epsfysize 3in\epsfbox{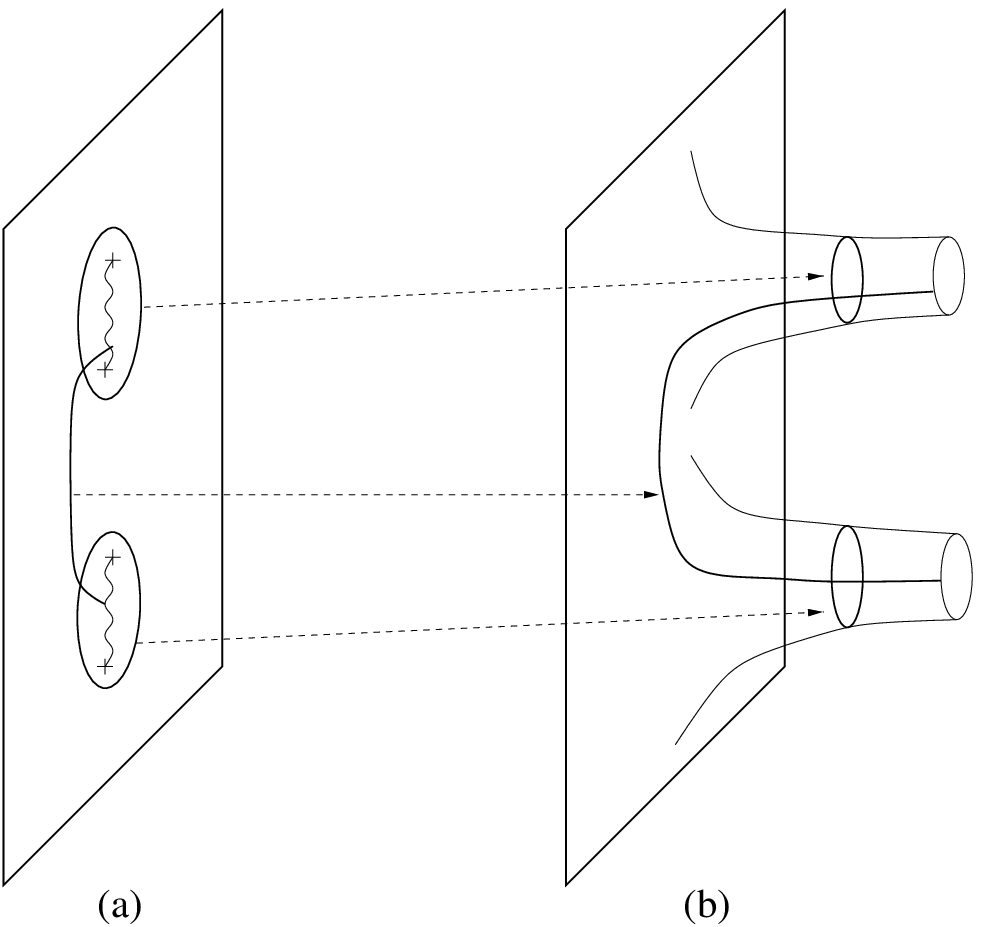}}

For the vector meson $W$, one now considers the surface
\eqn\Wsurface{
|t|= {1 \over 2},\qquad v=\bar v.
}
The pull-back of the K\"ahler form clearly vanishes, and it is also
easy to see that the pull-back of $\Omega$ is purely imaginary. It thus
saturates the conditions to be a supersymmetric surface. 
Can we also make it into a bounded surface? In fact, 
along the straight line segment on the $v$-plane between 
$v=\sqrt{\phi^2 + 1}$ and $v=\sqrt{\phi^2 - 1}$,
and also along the straight line segment between 
$v=-\sqrt{\phi^2 + 1}$ and $v=-\sqrt{\phi^2 - 1}$,
we have $|t| = {1 \over 2}$ on $\Sigma$.
As noted above, these are really closed
curves on $\Sigma$ winding around the two tubes that connect the upper
and lower sheets. Cutting off the surface \Wsurface\ along these
curves defines the vectormultiplet surface $D_W$, which thus has the
topology of a cylinder. An important fact is that the two boundary curves
are of opposite orientations on the upper sheet of $\Sigma$. 

For the monopole $M$, we take the surface\footnote{\dag}{If one is 
considering the case of purely imaginary $\phi$, the same kind of construction
gives the dyon instead.}
\eqn\Msurface{
t=\bar t, \qquad v=\bar v .
}
Again the BPS condition is satisfied trivially, and we only need to ask
if this surface contains a bounded component. In fact, both the upper and
the lower sheets of $\Sigma$ intersect with this surface along the 
straight line segments on the $v$-plane between 
$v = - \sqrt{\phi^2 - 1}$ and $v = \sqrt{\phi^2 + 1}$. Cutting off the
surface \Msurface\ along the closed curve formed by these segments
defines the monopole surface $D_M$, which thus has the topology of a
disc. Note that $D_M$ actually collapses to a point as
$\phi \rightarrow 1$. Of course, this simply reproduces the well-known
strong-coupling singularity on the moduli space of vacua where
the monopole becomes massless. The surfaces $D_W$ and $D_M$ are depicted
in figure 3 below.

\ifig\membranes{The BPS-saturated surfaces $D_W$ and $D_M$ with 
boundaries on $\Sigma$ corresponding to the vector meson and the
monopole.}{\epsfysize 2.5in\epsfbox{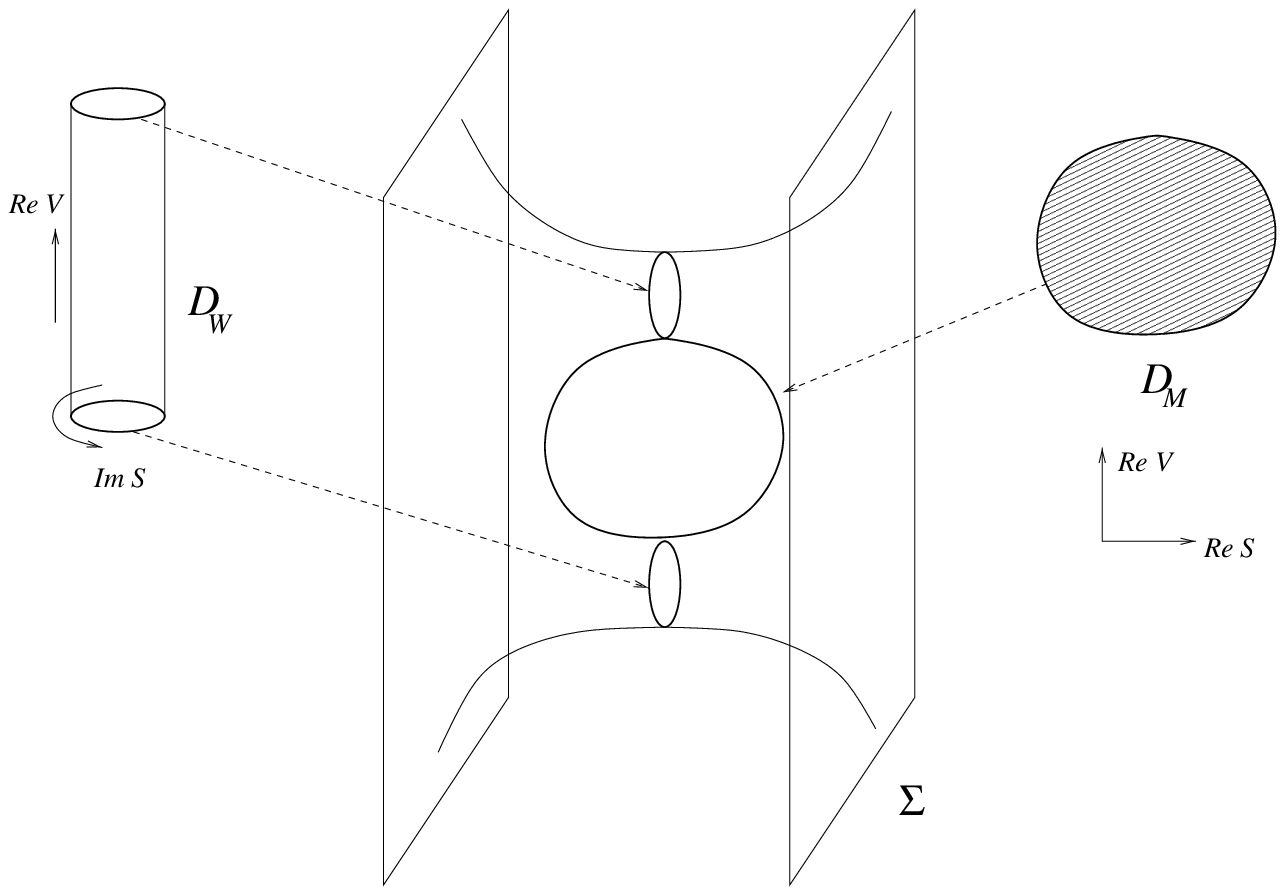}}

For the quark $Q$, we consider a situation where some $\phi_a$
is close to some $m_i$ and the others are far away. The reality condition
will be that we are at weak coupling, i.e. that 
${\rm Re \;} \log \Lambda \rightarrow - \infty$. For simplicity, we shift
and rescale $v$ so that $\phi_a = 1$ and $m_i = -1$. The upper sheet of
$\Sigma$ is then given by
\eqn\quarksigma{
s = \log {v + 1 \over v - 1} ,
}
where we have shifted $s$ by $\log \Lambda$. The quark surface $D_Q$ should be
holomorphically embedded with respect to the complex structure in which
\eqn\zwzero{
\eqalign{
z & = {1 \over 2} (s + \bar{s} + v - \bar{v}) \cr
w & = {1 \over 2} (v + \bar{v} - s + \bar{s})
}
}
are holomorphic variables. We note that $\Sigma$ is invariant under
the transformations 
$(s, v) \rightarrow (\bar{s}, \bar{v})$ and $(s, v) \rightarrow (-s, -v)$.
By symmetry, the same should hold for the quark surface $D_Q$, 
which must therefore be given by an equation of the form
\eqn\quarkf{
w = f(z) ,
}
where the holomorphic function $f$ is real for real argument and also odd,
i.e. $f(-z) = - f(z)$.
It follows that $D_Q$ cannot intersect the upper sheet of $\Sigma$ 
anywhere on the straight
line segment on the $v$-plane between $v = -1$ and $v = 1$, since 
here we have $z$ real but ${\rm Im \;} w = \pm \pi$.
Although we do not know the exact form of $f(z)$, it is not difficult to
numerically find a holomorphically embedded surface that comes close to having
a boundary on $\Sigma$ and thus approximates the true surface $D_Q$. 
To this end, we note that $f(z)$ can be expanded in a power series as
\eqn\powerseries{
f(z) = \sum_{k = 0}^\infty c_{2 k + 1} z^{2 k + 1} ,
}
where the coefficients $c_{2 k + 1}$ are real. To have a finite area
solution for $D_Q$, this series must converge everywhere on $D_Q$. We can
then approximate $f(z)$ by truncating the series after the first few terms.
In this way, one finds that the surface $D_Q$
is an almost flat disc bounded by an ellipse-like curve 
on the upper sheet of $\Sigma$ that surrounds,
but does not approach, the singular points at $v = -1$ and $v = 1$.  

\subsec{The topology of vectormultiplets and hypermultiplets}
We have discussed a few specific examples of supersymmetric two-branes 
as $N=2$ BPS-saturated states. An interesting feature is that the vector
meson, which is a vectormultiplet, is realized as a cylinder with two
boundaries, whereas the monopole and quark, which are hypermultiplets,
are realized as discs with a single boundary\footnote{*}{
We can also realize the matter sector by including six-branes in a
type IIA description or taking $Q^4$ to be a Taub-NUT space in the
$M$-theory description \Witten. The quark is then represented as a 
string stretching from a four-brane to a six-brane or as a surface
with a boundary on $\Sigma$ that winds around the compact direction
of $Q^4$. The other end of the surface is at a Taub-NUT center, where
the compact direction degenerates to a point, so this is a disc
rather than a cylinder.}.   
In fact, the
homology classes on the spectral curve that correspond to vector mesons
are represented by closed contours in opposite directions around two pairs
of branch points. This has to be the boundary of a cylinder or two separate
discs. However, the latter possibility would mean that each disc existed
separately as a BPS-saturated state, which is known not to be the case.
Also, all other known cases of BPS-saturated states, i.e. solitons and
quarks, have homology classes that can be represented by a single closed
contour that could be the boundary of a disc. These examples lead us to
formulate a conjecture: Surfaces with the topology of a cylinder and a 
disc correspond to, respectively, BPS-saturated vectormultiplets and 
hypermultiplets in the $N=2$ and $D=4$ theories. (Given the severe constraints
on supersymmetric surfaces with boundary on $\Sigma$, it seems unlikely
that such a surface could have more than two boundary components, except
at isolated points in the moduli space of $\Sigma$.) It would be interesting
to confirm this conjecture directly from the world-volume theory 
of the two-brane and the five-brane.

\subsec{The phenomenon of marginal stability}
By charge and energy conservation, BPS-saturated states are stable at a 
generic point in the moduli space of vacua. 
Furthermore, the spectrum of such states
is in general stable under small variations of the moduli. The only 
exception is when the quantum numbers of a BPS-saturate state is the sum 
of the quantum numbers of two other BPS-saturated states, and the central
charges of the three states have the same phase. As such a domain wall
of marginal stability in the moduli space of vacua is approached, 
the mass of the heaviest state approaches the sum of the masses of the 
lighter states. It is then {\it possible} for it to decay exactly at the 
domain 
wall and be absent from the spectrum on the other side. We emphasize that
these conditions are necessary but not sufficient for the decay of a
BPS-state.

The $M$-theory picture gives some further insight on this issue. The
stability of BPS-states away from their domain walls of marginal stability
means that given the two surfaces $\Sigma$ and
$D$, it must in general be possible to accommodate a small deformation
of $\Sigma$ by a deformation of $D$ so that $D$ still has its boundary 
on $\Sigma$. (Obviously we only consider deformations that preserve
the property that $\Sigma$ and $D$ are holomorphically embedded with 
respect to orthogonal complex structures.) 

For a hypermultiplet to decay
into two other hypermultiplets, the corresponding disc surface must
degenerate into two discs whose boundaries touch in a point.
This means that the intersection number of the corresponding homology 
classes on $\Sigma$ is $\pm 1$, or, equivalently,
that the symplectic product of the
electric-magnetic charge vectors equals $\pm 1$. These states are
thus mutually non-local. 
As an example, we take a dyonic soliton and a quark in $SU(2)$ 
Yang-Mills theory with a fundamental flavor. The corresponding
electric-magnetic charges $(q_e, q_m)$ and $(q_e^\prime, q_m^\prime)$
are $(2 n, 1)$ and $(1, 0)$
respectively, so that the symplectic product 
$q_m q_e^\prime - q_e q_m^\prime$ equals $1$.
In a certain domain in the moduli space, these states can form a bound
state 
\ref\Jackiw{
R. Jackiw and C. Rebbi, `Solitons with fermion number $1/2$',
{\it Phys. Rev.} {\bf D13} (1976) 3398.
}
that decays into its constituents as the corresponding domain wall
of marginal stability is crossed. More general situations of this kind were
studied in
\ref\Henningson{
M. Henningson, `Discontinuous BPS spectra in $N = 2$ gauge theory',
{\it Nucl. Phys.} {\bf B461} (1996) 101, {\tt hep-th/9510138}.
},
with the result that decay of hypermultiplet bound states is indeed 
related to the mutual non-locality of the constituents.

Similarly, for a vectormultiplet to decay into two hypermultiplets, the
corresponding cylinder surface must degenerate into two discs whose
boundaries touch in two points.
This means that the intersection number of the corresponding homology
classes on $\Sigma$ is $\pm 2$, so non-locality of the
constituents is again an essential requirement.
An example of this process can be found in pure $SU(2)$ gauge theory,
where the vector meson of electric-magnetic charge $(2, 0)$ can decay 
into two dyons with charges $(2 n, 1)$ and $(2 - 2 n, -1)$ so that
the symplectic product equals $2$.

The decay processes described in the last two paragraphs are fairly
easy to visualize. From a topological point of view, it seems much
less natural for a state to decay into two vectormultiplets or a
vectormultiplet and a hypermultiplet, at least at a generic point
along a domain wall of marginal stability. We would thus like to end
with the speculation that the only marginal stability decays that are
possible are those that we have described, namely that either a 
hypermultiplet or a vectormultiplet decays into two hypermultiplets. 
A proof of this statement or a counterexample
would be interesting.

\vskip 5 mm

We would like to thank Chalmers University of Technology (M.H.) and
the Theory Division at CERN and the Aspen Center for Physics (P.Y.)
for their hospitality. P.Y. is supported in part by U.S. Department of 
Energy.

\listrefs

\bye